\documentclass{optica-article}
\journal{oe}
\articletype{Research Article}
\begin{document}
\title{Generation of strong mechanical squeezing through the joint effect of two-tone driving and parametric pumping}
\author{Xiao-Jie Wu\authormark{1}, Huan-Huan Cheng\authormark{1}, Qiannan Wu\authormark{1,2,3}, Cheng-Hua Bai\authormark{1}, and Shao-Xiong Wu\authormark{1,*}}
\address{\authormark{1} School of Semiconductor and Physics, North University of China, Taiyuan 030051, China\\
\authormark{2} School of Instrument and Intelligent Future Technology, North University of China, Taiyuan 030051, China\\
\authormark{3} Academy for Advanced Interdisciplinary Research, North University of China, Taiyuan 030051, China}
\email{\authormark{*}sxwu@nuc.edu.cn}

\begin{abstract}
We propose an innovative scheme to efficiently prepare strong mechanical squeezing through utilizing the synergistic mechanism of two-tone driving and parametric pumping in an optomechanical system. By reasonable choosing the system parameters, the proposal highlights the following prominent advantages: the squeezing effect of the cavity field induced by the optical parametric amplifier can be transferred to the mechanical oscillator, which has been squeezed by the two-tone driving, and the degree of squeezing of the mechanical oscillator will surpass that obtained by any single mechanism; the joint mechanism can enhance the degree of squeezing significantly and break the 3 dB mechanical squeezing limit, which is particularly evident in range where the red/blue-detuned ratio is sub-optimal; the mechanical squeezing achieved through this distinctive joint mechanism exhibits notable robustness against both thermal noise and decay of mechanical oscillator. Our project offers a versatile and efficient approach for generating strong mechanical squeezing across a wide range of conditions.
\end{abstract}

\section{Introduction}
Cavity optomechanics \cite{Kippenberg2008, Favero2009} is an interdisciplinary frontier field that bridges quantum and classical mechanics, primarily focusing on the intricate interaction between light (optical fields) and mechanical motion within a resonant cavity \cite{Aspelmeyer2014}. Light can stimulate abundant dynamic behaviors in the interaction with matter, enabling precise control and manipulation of mechanical states via radiation pressure \cite{Ashkin1980}. The cavity optomechanical system can provide versatile platforms for various research fields, such as quantum entanglement \cite{Vitali2007, Zhou2011, Li2017, Jiao2020, Hu2020, Jiao2022, Lai2022, Wu2023, Xia2023, Wu2024}, photon blockade \cite{Liao2013, Wang2015, Li2019, WangDY2020, Yang2021, Wang2020, Gao2023}, mechanical squeezing \cite{Liao2011, Asjad2014, Bai2020, Xiongbiao2020}, non-reciprocal transmission \cite{Ruesink2016, Bernier2017, Peng2023, Wang2024, Ullah2024}, optomechanical cooling \cite{Teufel2011, Liu2015, Gan2019, Lau2020, Wen2022}, quantum synchronization \cite{Zhirov2006, Roulet2018, Sonar2018, Liao2019}, and so on.

The squeezed state, characterized by a variance in one quadrature component that falls below the standard quantum limit, is an essential resource in the field of quantum optics \cite{Walls1983,Scully1997}. The achievement of squeezing in mechanical resonator is a  milestone in the enhancement of measurement sensitivity in quantum information processing, especially in quantum metrology \cite{Caves1980, Braunstein2005, Peano2015} and gravitational wave detection \cite{Caves1981, Abbott2016, Barsotti2018}. Recently, significant progresses have been made in the generation and application of squeezed states within the realm of cavity optomechanics. Through extensive experimental and theoretical researches, scholars have demonstrated a variety of methods for achieving mechanical oscillator squeezing \cite{Naeini2013, Pirkkalainen2015, Wollman2015, Lei2016, Lin2024} under various driving conditions. The achieved mechanical squeezing is relatively weak and does not exceed the 3 dB squeezing limit when only parametric pumping \cite{Agarwal2016, Bai2019} or periodic modulation of the external driving amplitude \cite{Liao2011, Mari2009, Gu2013, Han2019, Guo2023} is applied. Aiming to break the 3 dB squeezing limit, which corresponds to a reduction of quantum noise fluctuations by half, numerous advanced strategies for strong mechanical squeezing have been proposed, such as two-tone driving \cite{Kronwald2013, Huang2021, Zhang2021, Zhao2024}, Duffing nonlinearity \cite{Lv2015, Xiong2020, Zhang2020, Bai2019} and others. All of these schemes can successfully incorporate the quantum interference effect \cite{Scully1997, Lei2023} into the motion of the mechanical oscillator, enabling the suppression of quantum noise components and achieving a remarkable squeezing effect.

Both parametric pumping \cite{Milburn1981, Wu1986, Agarwal2016, Qin2018, Huang2020, Huang2023} and two-tone driving \cite{Kronwald2013, Huang2021, Zhao2024, Zhang2021} serve as pivotal techniques for generating mechanical squeezed states in an optomechanical system. In the scheme of mechanical squeezing via parametric amplification \cite{Agarwal2016}, the achievable squeezing is inherently limited to below 3 dB, which is primarily constrained by the stability of the system. However, a squeezed steady state of mechanical oscillator that overcomes 3 dB limit can be prepared by combining parametric pumping with Duffing nonlinearity \cite{Bai2019}, both of which individually yield squeezing below 3 dB. Ref. \cite{Kronwald2013} reports a method to generate a mechanical squeezed steady state that surpasses the 3 dB limit using two control lasers with different amplitudes (the powers of red-detuned field and blue-detuned field fall within a certain range). The quantum squeezing engineered by two-tone driving was substantiated through experiment, for example, Ref. \cite{Pirkkalainen2015} realized that a micromechanical resonator can be placed in a squeezed quantum state in microwave optomechanical systems using two-tone driving techniques; Ref. \cite{Lei2016} successfully prepared a quantum squeezed state that exceeds the traditional 3 dB squeezing limit. Nevertheless, the mechanical squeezing remains relatively modest, with a level less than 3 dB, when the amplitude ratio of the red/blue-detuned laser is far from optimal range.  A natural question arises: whether the mechanical squeezing induced by two-tone driving can be further enhanced through an auxiliary method, such as parametric pumping, thereby surpassing the 3 dB limit in most range?

Inspired by the aforementioned work, we propose a novel proposal in this paper, which generates mechanical squeezing by simultaneously harnessing the effects of two-tone driving and parametric pumping. Our scheme aims to utilize the joint effect of two-tone driving and parametric pumping to overcome the limitation that two-tone driving alone cannot break through the 3 dB limit under the sub-optimal condition. Firstly, we find that the time-dependent terms have a negligible influence on the system's dynamical evolution. Therefore, the standard linearization method can be employed to analyze the properties of the steady-state squeezing. Secondly, we explore the enhancement of mechanical squeezing and highlight the crucial role of blue-detuned driving in this proposal. Specifically, we demonstrate how the squeezing of the cavity field, generated by an optical parametric amplifier, is efficiently transferred to the mechanical oscillator by analyzing Wigner function. Furthermore, we conduct a thorough investigation into the influence of various system parameters on the mechanical squeezing, which reveals how alterations in these parameters affect the degree of mechanical squeezing. The advantage of our scheme lies in its capability to attain strong squeezing up to 19.49 dB through the joint effect; especially, in the range that the amplitude ratio of the red/blue-detuned laser is far from optimal, the degree of mechanical squeezing will still be greater than 3 dB.

The paper is structured as follows. In Sec. \ref{sec2}, the system model and the Hamiltonian are introduced, and the steady-state mean values for both the cavity field and the mechanical oscillator are calculated. In Sec. \ref{sec3}, we derive the quantum Langevin equations for the quadrature fluctuation operators and obtain the dynamical equations for the covariance matrix. In Sec. \ref{sec4}, we show the mechanism by which squeezing is transferred from the cavity field to the mechanical oscillator, and the robustness of the mechanical squeezing under different conditions is also verified. The analytical solutions of the mechanical squeezing are presented in Sec. \ref{sec5}, and the conclusion is given at the end.

\section{Theoretical model}\label{sec2}
\begin{figure}[t]
\centering
\includegraphics[width=0.5\columnwidth]{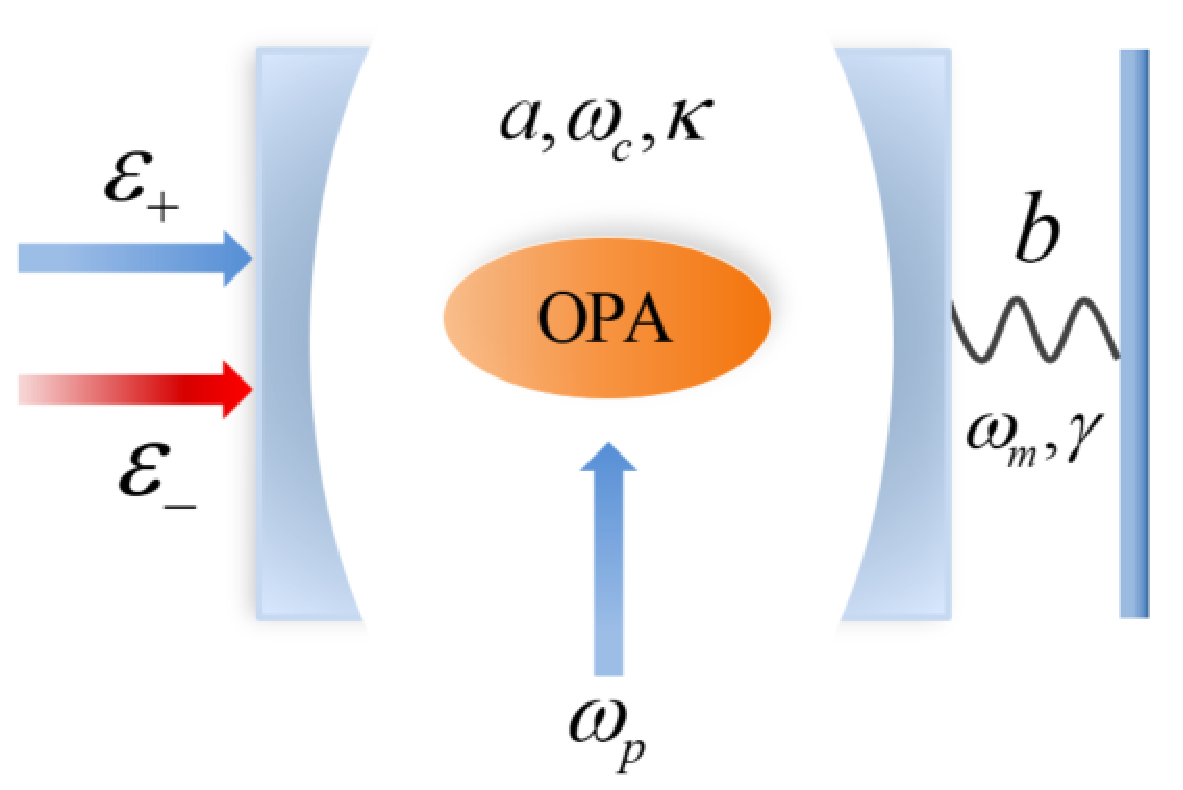}
\caption{Schematic diagram of the optomechanical system. The system consists of a mechanical oscillator coupled to an optical cavity, and the optical cavity is driven by two-tone detuned lasers: one is red-detuned and the other is blue-detuned. A degenerate second-order nonlinear crystal OPA is trapped within the cavity and is pumped by a coherent parametric laser.}\label{fig:1}
\end{figure}

As schematically illustrated in Fig. \ref{fig:1}, we consider an optomechanical system that consists of a fixed mirror, a movable mirror (acting as a mechanical oscillator), and a degenerate optical parametric amplifier (OPA) embedded within the optical cavity. The optical cavity is driven by two-tone lasers: one is red-detuned and the other is blue-detuned, and the degenerate OPA is pumped by a coherent parametric laser with frequency $\omega_p$. We will investigate the combined effect of the two-tone driving and the parametric pumping on the squeezing of the mechanical oscillator. The Hamiltonian of the whole system is described as follows (in units of $\hbar$):
\begin{align}
H=&\omega_c{a^\dag}a+\omega_m{b^\dag}b-g_0{a^\dag}a(b^\dag+b) +(\varepsilon_+e^{-i\omega_+t}+\varepsilon_-e^{-i\omega_-t})a^\dag +(\varepsilon_+e^{i\omega_+t}+\varepsilon_-e^{i\omega_-t})a\notag\\ &+\frac{iG}{2}(e^{i\theta}a^{\dag2}e^{-i\omega_pt}-e^{-i\theta}a^2e^{i\omega_pt}).\label{eq:H}
\end{align}
The first term represents the free Hamiltonian of the optical cavity with frequency $\omega_c$, which is accompanied by a decay rate $\kappa$. Here, $a$ and $a^\dag$ denote annihilation and creation operators, and meet the canonical commutation relation $[a,a^\dag]=1$. The second term describes the free Hamiltonian of the mechanical oscillator with frequency $\omega_m$ and decay rate $\gamma_m$, and $b$ and $b^\dag$ are annihilation and creation operators of the mechanical oscillator. The third term expresses the interaction Hamiltonian between the cavity field and the mechanical mode with single photon coupling strength $g_0$. The fourth and fifth terms are Hamiltonian of the two-tone driving that applied to the cavity field, where $\varepsilon_{\pm}$ is the blue/red-detuned driving strength governed by $\varepsilon_{\pm}= \sqrt{\kappa P_{\pm}/\hbar\omega_{\pm}}$ under the driving power $P_{\pm}$ with frequency $\omega_{\pm}=\omega_c\pm\omega_m$. The final term means that the degenerate OPA is pumped by a coherent parametric laser, where $G$ signifies the gain coefficient of OPA and $\theta$ is the relative phase of the pumping laser.

Incorporating the dissipative dynamics and the inherent noise, the evolution of the system can be expressed by the quantum Langevin equations \cite{Scully1997}:
\begin{align}
\dot{a}=&-(i\omega_c+\frac{\kappa}{2})a+ig_0a(b^\dag+b)+Ge^{i\theta}e^{-i\omega_pt}a^\dag -i(\varepsilon_+e^{-i\omega_+t}+\varepsilon_-e^{-i\omega_-t})+\sqrt{\kappa}a_{\text{in}},\notag\\ \dot{b}=&-(i\omega_m+\frac{\gamma_m}{2})b+ig_0{a^\dag}a+\sqrt{\gamma_m}b_{\text{in}}.
\label{eq:QLEs}
\end{align}
In the above system of equations, $a_\text{in}$ denotes the cavity field's zero-mean input vacuum noise operator, and $b_\text{in}$ means the thermal noise operator associated with the mechanical oscillator. These operators can be characterized by non-zero noise correlation functions, which are given by $\langle a_\text{in}(t)a_{\text{in}}^{\dag}(t')\rangle=\delta(t-t')$, $\langle a_\text{in}^{\dag}(t)a_{\text{in}}(t')\rangle=0$ and $\langle b_\text{in}(t)b_{\text{in}}^{\dag}(t')\rangle= (n_m^{\text{th}}+1)\delta(t-t')$, $\langle b_\text{in}^{\dag}(t)b_{\text{in}}(t')\rangle= n_m^{\text{th}}\delta(t-t')$. Here, $n_m^{\text{th}}=[\exp(\hbar\omega_m/k_BT)-1)]^{-1}$ represents the mean phonon number and is intimately connected to the thermal environmental temperature $T$, where $k_B$ is the Boltzmann constant.

In order to evaluate the time-dependent terms induced by the frequency difference between the two-tone driving lasers, we will introduce coefficient symbols $\alpha_+$ and $\alpha_-$ to rewrite the steady-state mean value of the cavity field:
\begin{align}
\alpha(t)=&\alpha_+e^{-i\omega_+t}+\alpha_-e^{-i\omega_-t},\label{eq:alpha}
\end{align}
where $\alpha_+$ and $\alpha_-$ denote the amplitudes of steady state that correspond to the positive (blue-detuned) and negative (red-detuned) frequency components of the cavity field, respectively, and their analytical expressions are derived as follows:
\begin{align}
\alpha_+=\frac{2\varepsilon_+(i\kappa+2\tilde{\omega}_c-2\omega_+)}{4G^2-\kappa^2-4(\omega_+-\tilde{\omega}_c)^2},\ \alpha_-=\frac{2\varepsilon_-(i\kappa+2\tilde{\omega}_c-2\omega_-)}{4G^2-\kappa^2-4(\omega_--\tilde{\omega}_c)^2}, \label{eq:alpha+-}
\end{align}
where the high frequency terms are reasonably omitted. In the resolved sideband range, for simplicity and without loss of generality, we can assume that the steady-state amplitude $\alpha_{\pm}$ is a real number and proportional to the driving strength $\varepsilon_{\pm}$ since the cavity decay rate is satisfied by the relation $\kappa\ll\omega_m$. In Eq. (\ref{eq:alpha+-}), the effective frequency $\tilde{\omega}_c$ is determined by $\tilde{\omega}_c=\omega_c-g_0(\beta^*+\beta)$, where $\beta$ denotes the steady-state amplitude of the mechanical oscillator. The relationship between the amplitudes of the cavity field and the mechanical oscillator satisfies
\begin{align}
\omega_m\beta-g_0|\alpha|^2=0.\label{eq:beta}
\end{align}
The decay rate of the mechanical oscillator is significantly lower than that of the cavity field, i.e., $\gamma_m\ll\kappa$, therefore, the influence of $\gamma_m$ will be sufficiently minor and hardly contribute to the dynamical evolution, and the corresponding term can be reasonably ignored in Eq. (\ref{eq:beta}).

\section{Quantum Fluctuations}\label{sec3}
Under the assumption of strong driving condition, both components of the steady-state average value of the cavity field, $|\alpha_+|^2$ and $|\alpha_-|^2$, will significantly exceed unity, so the standard linearization method will be employed to handle the nonlinear quantum Langevin equations (\ref{eq:QLEs}). We can reformulate the operators by utilizing the displacement operators $a=\alpha+\delta a$ and $b=\beta+\delta b$, which are rewritten as the sum of the steady-state mean values and the corresponding fluctuation operators. Consequently, the linearized quantum Langevin equations of the fluctuation operators $\delta a$ and $\delta b$ can be derived as follows:
\begin{align}
\delta\dot{a}=&-(i\tilde{\omega}_c+\frac{\kappa}{2})\delta a+ig(\delta{b^\dag}+\delta{b}) +Ge^{i\theta}e^{-i\omega_pt}\delta{a^\dag}+\sqrt{\kappa}a_{\text{in}},\notag\\ \delta\dot{b}=&-(i\omega_m+\frac{\gamma_m}{2})\delta b+i(g^*\delta a+g\delta{a^\dag}) +\sqrt{\gamma_m}b_{\text{in}},
\label{eq:zhangluo}
\end{align}
where the coupling coefficient satisfies $g=g_0\alpha$. To seek solving the time-dependent system of differential equations (\ref{eq:zhangluo}), we will introduce the slow-varying operators $\delta a=\delta\tilde{a}e^{-i\tilde{\omega}_ct}$, $\delta b=\delta\tilde{b}e^{-i\omega_mt}$, $a_{\text{in}}=\tilde{a} _{\text{in}}e^{-i\tilde{\omega}_ct}$ and $b_{\text{in}}=\tilde{b}_{\text{in}}e^{-i\omega_mt}$. The motions of the fluctuation operators $\delta\tilde{a}$ and $\delta\tilde{b}$ can be obtained through the following:
\begin{align}
\delta\dot{\tilde{a}}=&i[f_2(t)\delta\tilde{b}^\dag+f_3(t)\delta\tilde{b}] +Ge^{i\theta}\delta\tilde{a}^\dag-\frac{\kappa}{2}\delta\tilde{a}+\sqrt{\kappa}\tilde{a}_{\text{in}},\notag\\ \delta\dot{\tilde{b}}=&i[f_1(t)\delta\tilde{a}+f_2(t)\delta\tilde{a}^\dag] -\frac{\gamma_m}{2}\delta\tilde{b}+\sqrt{\gamma_m}\tilde{b}_{\text{in}}, \label{eq:youxiaozhangluo}
\end{align}
where $f_1(t)=g_-+g_+e^{2i\omega_mt}$, $f_2(t)=g_++g_-e^{2i\omega_mt}$ and $f_3(t)=g_-+g_+e^{-2i\omega_mt}$ with $g_{\pm}=g_0\alpha_{\pm}$. Under the weak coupling limit and the rotating wave approximation (RWA), the pumping frequency is judiciously chosen to satisfy the frequency matching condition $\omega_p=2\tilde{\omega}_c$.

To facilitate investigating the squeezing characteristics of the mechanical oscillator, it is advantageous to introduce the position and momentum quadrature fluctuation operators and the quadrature noise operators inherent to the cavity field, which can be written as
\begin{align}
\delta{X}=\frac{\delta\tilde{a}+\delta\tilde{a}^\dag}{\sqrt{2}},&\ \delta{Y}=\frac{\delta\tilde{a}-\delta\tilde{a}^\dag}{i\sqrt{2}},\notag\\ X_{\text{in}}=\frac{\tilde{a}_{\text{in}}+\tilde{a}^\dag_{\text{in}}}{\sqrt{2}},&\ Y_{\text{in}}=\frac{\tilde{a}_{\text{in}}-\tilde{a}^\dag_{\text{in}}}{i\sqrt{2}}. \label{eq:azhengjiao}
\end{align}
Meanwhile, the position and momentum quadrature operators of the mechanical oscillator and the corresponding quadrature noise operators can be expressed as
\begin{align}
\delta{Q}=\frac{\delta\tilde{b}+\delta\tilde{b}^\dag}{\sqrt{2}},&\ \delta{P}=\frac{\delta\tilde{b}-\delta\tilde{b}^\dag}{i\sqrt{2}},\notag\\
Q_{\text{in}}=\frac{\tilde{b}_{\text{in}}+\tilde{b}^\dag_{\text{in}}}{\sqrt{2}},&\ P_{\text{in}}=\frac{\tilde{b}_{\text{in}}-\tilde{b}^\dag_{\text{in}}}{i\sqrt{2}}.\label{eq:bzhengjiao}
\end{align}

Based on the linearized quantum Langevin equations (\ref{eq:youxiaozhangluo}), the dynamics of the cavity field and the mechanical oscillator can be rewritten succinctly as a matrix representation
\begin{align}
\dot{\mathcal{U}}(t)=\mathcal{M}(t)\mathcal{U}(t)+\mathcal{N}(t).\label{eq:juzhen}
\end{align}
In Eq. (\ref{eq:juzhen}), the vector $\mathcal{U}(t)=[\delta{X},\delta{Y}, \delta{Q},\delta{P}]^T$ is the system's quadrature fluctuation operators, the vector $\mathcal{N}(t)=[\sqrt{\kappa}{X}_{\text{in}}, \sqrt{\kappa}{Y}_{\text{in}}, \sqrt{\gamma_m}{Q}_{\text{in}}, \sqrt{\gamma_m}{P}_{\text{in}}]^T$ is associated with the noise processes, and the drift matrix $\mathcal{M}(t)$ that governs the dynamical dissipation process occurring  between the cavity field and the mechanical mode can be expressed as a $4\times4$ matrix form
\begin{align}
\mathcal{M}(t)= \left[
  \begin{array}{cccc}
    G\cos\theta-\frac{\kappa}{2}  & G\sin\theta                     & -I(f_{23}^+)            & R(f_{23}^-)\\
         G\sin\theta              & -G\cos\theta-\frac{\kappa}{2}   & R(f_{23}^+)             & I(f_{23}^-)\\
          -I(f_{12}^+)            & R(f_{21}^-)                     & -\frac{\gamma_m}{2}     & 0\\
           R(f_{12}^+)            & I(f_{21}^-)                     & 0                       & -\frac{\gamma_m}{2}
\end{array}\right],\label{eq:piaoyijuzhen}
\end{align}
where $R(f)$ and $I(f)$ represent the real and imaginary parts of the complex coupling coefficient $f_{jk}^{\pm}=f_j(t){\pm}f_k(t)$, respectively. According to the Routh-Hurwitz criterion \cite{DeJesus1987}, the system is stable if and only if all the eigenvalues of the drift matrix $\mathcal{M}(t)$ have negative real parts. Through a direct but tedious calculation, the Routh-Hurwitz criterion can be simplified as the following inequalities
\begin{align}
&\frac{\gamma_m^2}{16}(\kappa^2-4G^2)+\frac{\kappa\gamma_m}{2}(g_-^2-g_+^2)+(g_-^2-g_+^2)^2>0,\notag\\
&\frac{\kappa}{4}(\kappa^2-4G^2)+[(g_-^2-g_+^2)+\frac{1}{4}(\gamma_m^2+3\kappa\gamma_m)]
(\kappa+\gamma_m)+\frac{1}{4}\kappa^2\gamma_m>0,\label{eq:R-H}\\
&\frac{1}{16}[\kappa\gamma_m(\kappa^2-4G^2)+4(g_-^2-g_+^2)(\kappa+\gamma_m)^2+\kappa\gamma_m^3+2\kappa^2\gamma_m^2]
[(\kappa^2-4G^2)+\gamma_m^2+2\kappa\gamma_m]>0.\notag
\end{align}
Obviously, the stability of the system is independent of the relative phase $\theta$. Taking into account the fact that $\gamma_m\ll\kappa$, a sufficient condition for the solution of Eq. (\ref{eq:R-H}) is that both $G<0.5\kappa$ and $g_+<g_-$ hold true, and this restriction is adopted in the analysis of the mechanical squeezing.

Due to the linearized dynamics of optomechanical system and the Gaussian nature of environmental noise, the evolution of system's steady state will converge toward a Gaussian state \cite{Weedbrook2012}. Consequently, the fluctuation dynamics of the system can be fully characterized by a $4 \times 4$ covariance matrix $\mathcal{V}(t)$ with elements $\mathcal{V}_{jk}\delta(t-t')= \langle\mathcal{U}_j(t)\mathcal{U}_k(t')+\mathcal{U}_k(t')\mathcal{U}_j(t)\rangle /2$, and  governed by the motion equation
\begin{align}
\frac{d\mathcal{V}(t)}{dt}=\mathcal{M}(t)\mathcal{V}(t)+\mathcal{V}(t)\mathcal{M}^T(t)+\mathcal{D}. \label{eq:dotV}
\end{align}
The diffusion matrix $\mathcal{D}$ is related to the noise correlation function, and its elements are defined as $\mathcal{D}_{jk}\delta(t-t') =\langle\mathcal{N}_j(t)\mathcal{N}_k(t')+ \mathcal{N}_k(t')\mathcal{N}_j(t)\rangle/2$. In our system, the diffusion matrix $\mathcal{D}$ can be simplified to a diagonal matrix
\begin{align}
\mathcal{D}=\text{diag}\left[\frac{\kappa}{2},\frac{\kappa}{2},\frac{\gamma_m}{2}(2n_m^{\text{th}}+1), \frac{\gamma_m}{2}(2n_m^{\text{th}}+1)\right].\label{eq:D}
\end{align}

In the following contents, we will explore the squeezing effect of the mechanical oscillator. Taking the position quadrature fluctuation operators as an example to demonstrate how the degree of squeezing is quantified, it is typically measured in units of dB and can be expressed as follows \cite{Walls1983, Scully1997}:
\begin{align}
S=&-10\log_{10}\left[\frac{\langle\delta Q(t)^2\rangle}{\langle\delta Q(t)^2\rangle_{\text{vac}}}\right].\label{eq:S}
\end{align}
The positional fluctuation of the mechanical oscillator in vacuum is denoted by $\langle \delta Q(t)^2\rangle_{\text{vac}}=1/2$, while $\langle \delta Q(t)^2\rangle$ represents the corresponding variance of the position operator in the covariance matrix $\mathcal{V}(t)$ since the mean value of the position operator is zero. Traditionally, when the degree of squeezing of a mechanical oscillator exceeds 3 dB, it is called strong squeezing.

\begin{figure}[b]
\centering
\includegraphics[width=0.45\columnwidth]{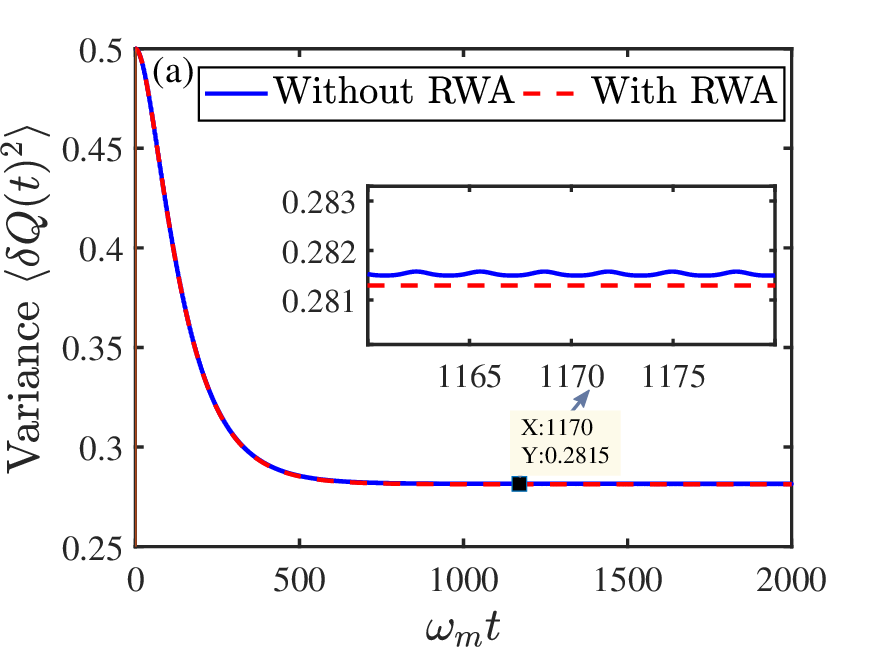}
\includegraphics[width=0.45\columnwidth]{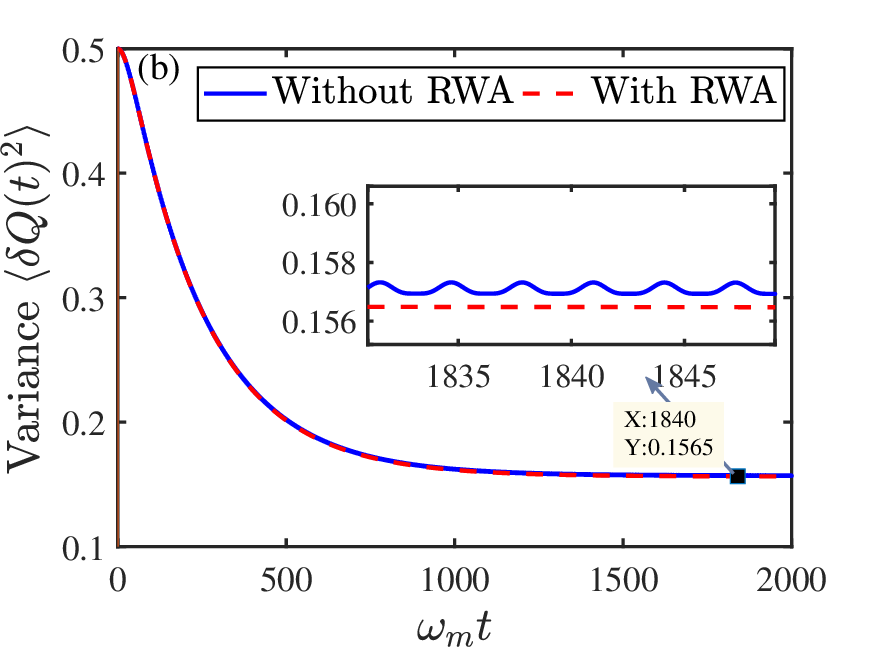}
\caption{The evolution of the mechanical oscillator's variance of the position operator without the RWA (blue solid line) and with the RWA (red dashed line). The gain coefficient is $G=0$ (a) and $G=0.4\kappa$ (b), and the other system parameters are set: $g_-=0.01\omega_m,g_+=0.0028\omega_m,n_m^{\text{th}}=0$ and $\theta=0$.}\label{fig:2}
\end{figure}

\section{Squeezing dynamics of the mechanical oscillator}\label{sec4}
In this section, we will initially focus on the contributions of the time-dependent terms $R(f)$ and $I(f)$ to Eq. (\ref{eq:piaoyijuzhen}). Under the conditions of weak coupling limit, i.e., $g_+,g_-\ll\omega_m$, the time-dependent terms can be regarded as high frequency oscillating terms and can be approximated as negligible, which do not affect the properties of the squeezing. Based on the contemporary experimental technique \cite{Li2009}: the frequency of the mechanical oscillator is assumed as $\omega_m=2\pi\times25.45 \text{MHZ}$, the decay rates for the system are set as $\kappa=0.05\omega_m$ and $\gamma_m=10^{-6}\omega_m$ and the frequency of the red-detuned laser was determined by the formula $\omega_-=2\pi c/\lambda_-$ with the laser's wavelength $\lambda_-=1564.25\text{nm}$. In Fig. \ref{fig:2}, we present the evolution of the variance of the position operator of the mechanical oscillator by numerical simulation under both the RWA and non-RWA conditions. The evolution trends with both treatments are exactly identical, reflecting the validity of the RWA approximation. So, it is reasonable to proceed with further analysis of the mechanical squeezing properties under the RWA condition. Even though, there exist oscillatory trajectories in the cases that RWA approximation are not applied, as zoomed in the insets of both Figs. \ref{fig:2}(a) and \ref{fig:2}(b). Fig. \ref{fig:2}(a) compares the variance of the positional fluctuation in the scenarios with and without the application of RWA, specifically when only the two-tone driving is present under the ratio of $g_+/g_-=0.28$. The thermal phonon environment is vacuum, and the relative phase is set as $\theta=0$. The single-photon coupling strength is defined as $g_0=2.0\times10^{-5}\omega_m$, and the coefficient representing the red-detuned laser is chosen as $g_-=0.01\omega_m$, which corresponds to the amplitudes of the red-detuned driving laser is $\varepsilon_-=5.1\times10^2\omega_m$ under the driving power $P_-\simeq0.1\text{mW}$. When the system attains a steady state, the minimal variance will reach $\langle \delta Q(t)^2\rangle=0.2815$. According to the criterion of squeezing, the fluctuation of position operator of the mechanical oscillator should be smaller than 1/2. It indicates that the mechanical mode had been squeezed, and the mechanism of squeezing stems from the interplay between the mechanical oscillator and the cavity field under the competition and cooperation effects of  the driving frequencies between the two-tone lasers. Specifically, the blue-detuned laser enhances the energy of the oscillator, whereas the red-detuned laser exerts a cooling effect on it \cite{Aspelmeyer2014,Kronwald2013}. In Fig. \ref{fig:2}(b), we delve deeper into evaluating  the joint effect of two-tone driving and parametric pumping on the mechanical oscillator, where the gain coefficient of the degenerate OPA is $G=0.4\kappa$. When the system reaches a steady state, the minimal fluctuation of the position operator will be further suppressed and attain $\langle\delta Q(t)^2\rangle=0.1565$. The physical mechanism underlying the results can be explained by the fact that a degenerate OPA can induce the squeezing of the cavity field, and the squeezing is subsequently transferred to the mechanical oscillator. Consequently, the variance of position operator of the mechanical oscillator is further suppressed by an auxiliary squeezing action.

\begin{figure}[b]
\centering
\includegraphics[width=0.45\columnwidth]{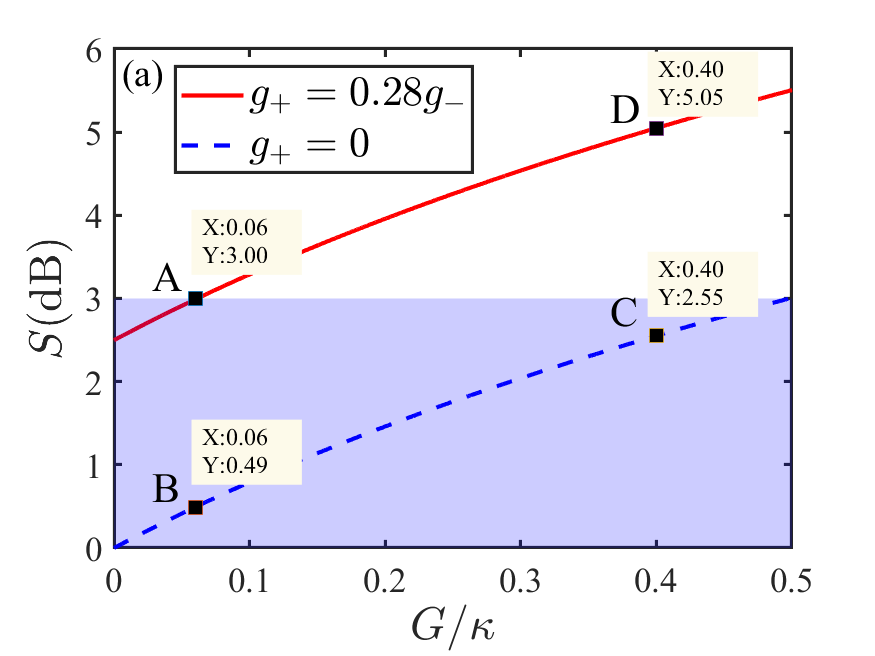}
\includegraphics[width=0.45\columnwidth]{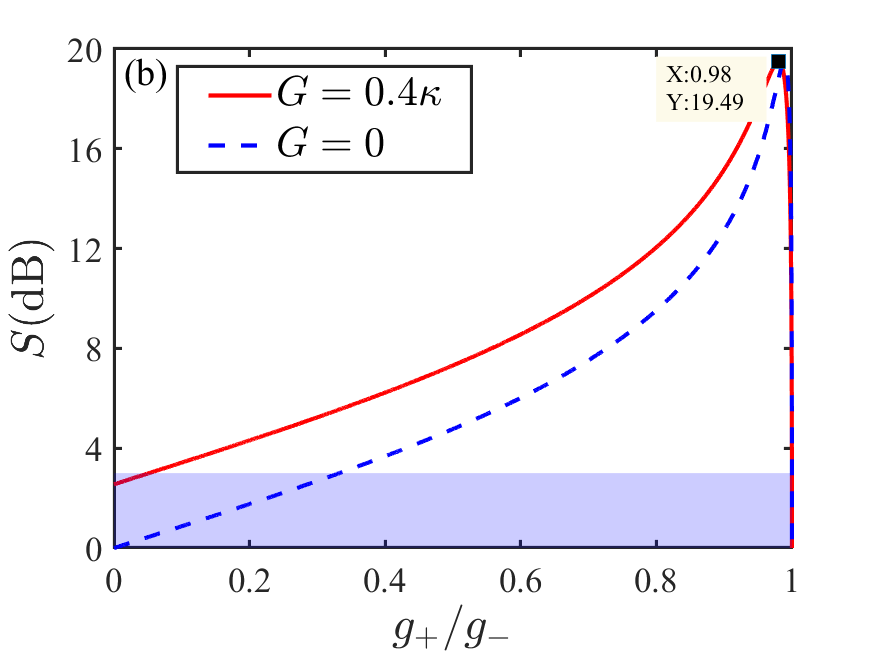}
\caption{The degree of mechanical squeezing as a function of the gain coefficient $G$ (a) and as a function of the $g_+/g_-$ ratio (b). In panel (a), the solid red line corresponds to $g_+=0.28g_-$ and the dashed blue line corresponds to $g_+=0$. In panel (b), the red solid line denotes $G=0.4\kappa$ and the blue dashed line denotes $G=0$. The purple shaded regions represent the areas where the degree of squeezing is below the 3 dB limit, and other parameters are the same as those in Fig. \ref{fig:2}.}\label{fig:3}
\end{figure}

In Fig. \ref{fig:3}(a), we numerically investigate how the degree of squeezing varies with the gain coefficient $G$ under two distinct conditions that only the red-detuned laser is present ($g_+=0$, the blue dashed line), and both the red-detuned and blue-detuned lasers are injected into the cavity ($g_+/g_-=0.28$, the red solid line). The parameter sets $\{g_+/g-_,G/\kappa,S/\text{dB}\}$ corresponding to the labeled points A, B, C, and D are chosen as $\{0.28, 0.06, 3.00\}$, $\{0,0.06,0.49\}$, $\{0,0.40,2.55\}$ and $\{0.28,0.40,5.05\}$, respectively. The point A reflects that the mechanical oscillator achieves strong squeezing of 3 dB under the joint effect of two-tone driving and parametric pumping. However, at point B, when the same parametric pumping is applied but without the blue-detuned laser, the mechanical oscillator exhibits a reduction in the degree of squeezing, which is only 0.49 dB. For the point C, although the mechanical squeezing is enhanced as the increasing of the gain coefficient $G$, it still fails to break the 3 dB squeezing limit. When both two-tone driving and parametric pumping are simultaneously employed, just as shown at point D, there is a substantial enhancement of the squeezing, and the 3 dB squeezing limit can be readily broken. In Fig. \ref{fig:3}(b), we discuss the role of the blue-detuned laser in the generation of mechanical squeezing. The blue dashed line is the scenario where the parametric pumping is absent, while the red solid line presents the case where the gain coefficient is set to $G=0.4\kappa$. We can observe that the degree of squeezing initially enhances and subsequently diminishes when the red/blue-detuned ratio $g_+/g_-$ is increased, and the peak of the degree of squeezing aligns with the optimal ratio of $g_+/g_-$. However, in the case that both identical blue-detuned driving and parametric pumping with gain coefficient $G=0.4\kappa$ are employed, the mechanical squeezing will be stronger (the maximum degree of squeezing is about 19.49 dB) than that achieved only by two-tone driving mechanism in most range. This indicates a crucial role of the blue-detuned laser in preparing mechanical squeezing, and the auxiliary application of a degenerate OPA can further enhance the squeezing of the mechanical oscillator.

\begin{figure}[t]
\centering
\includegraphics[width=0.3\columnwidth]{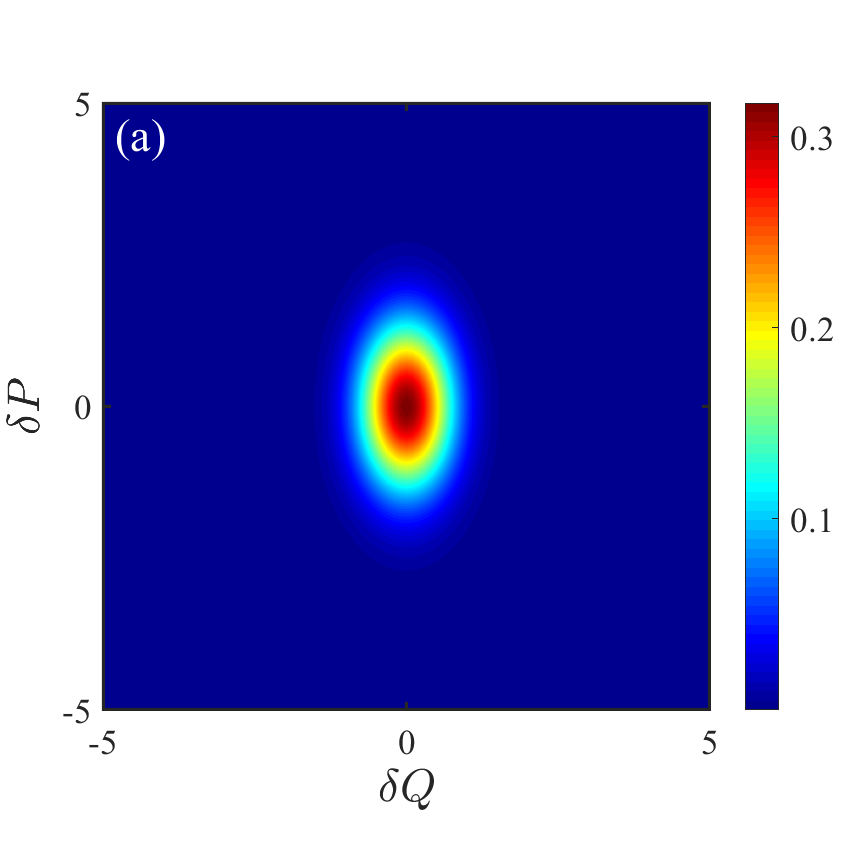}
\includegraphics[width=0.3\columnwidth]{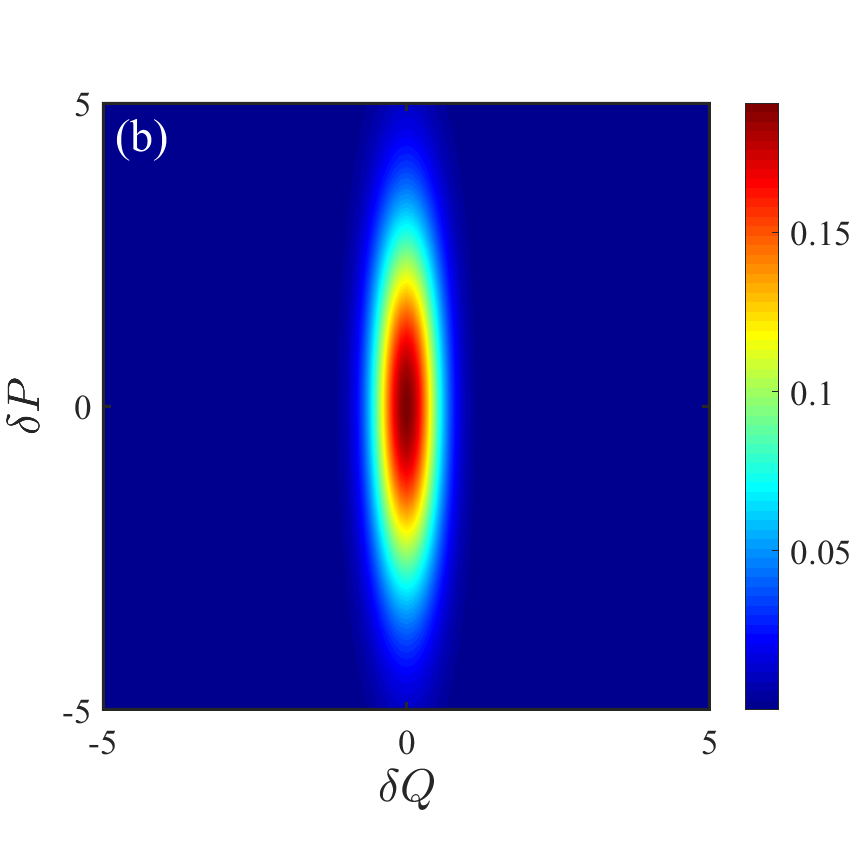}
\includegraphics[width=0.3\columnwidth]{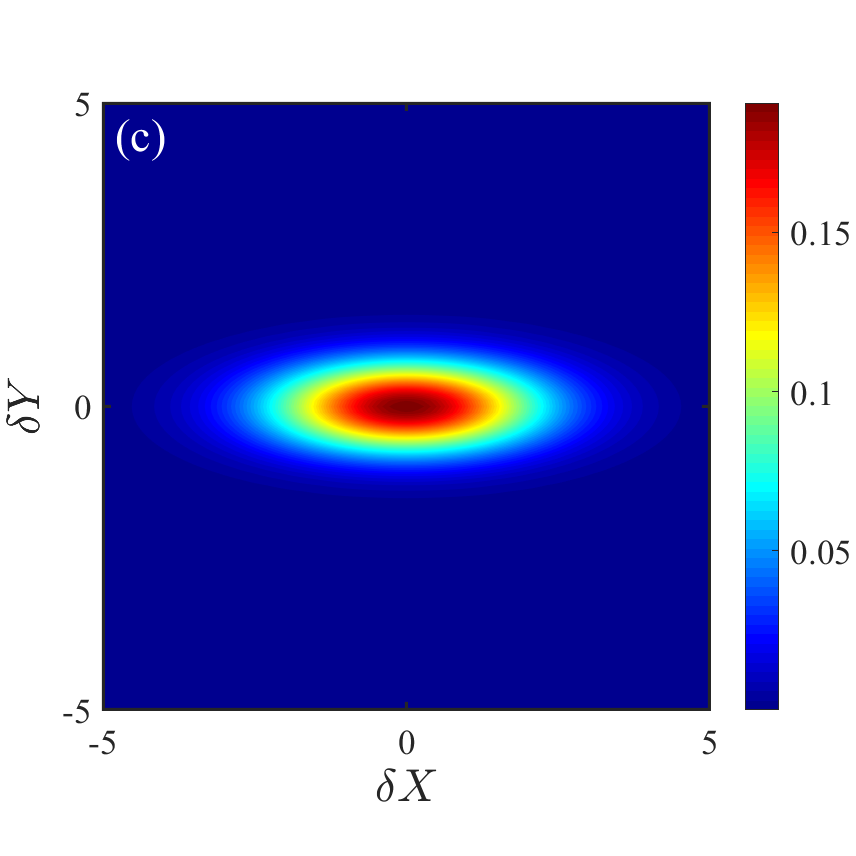}
\includegraphics[width=0.3\columnwidth]{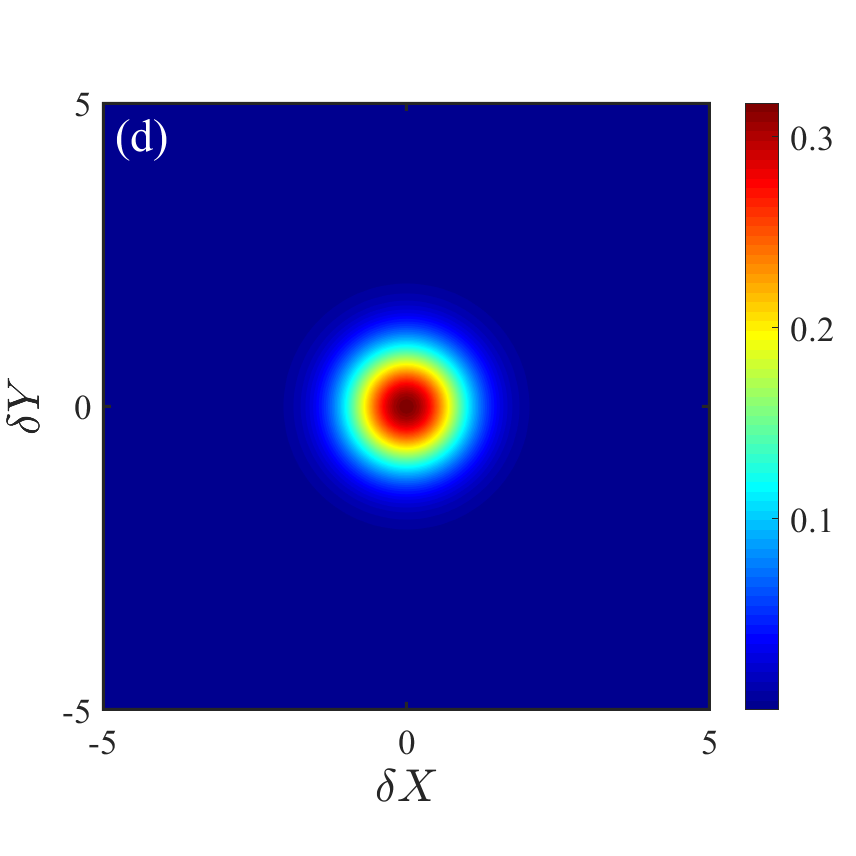}
\includegraphics[width=0.3\columnwidth]{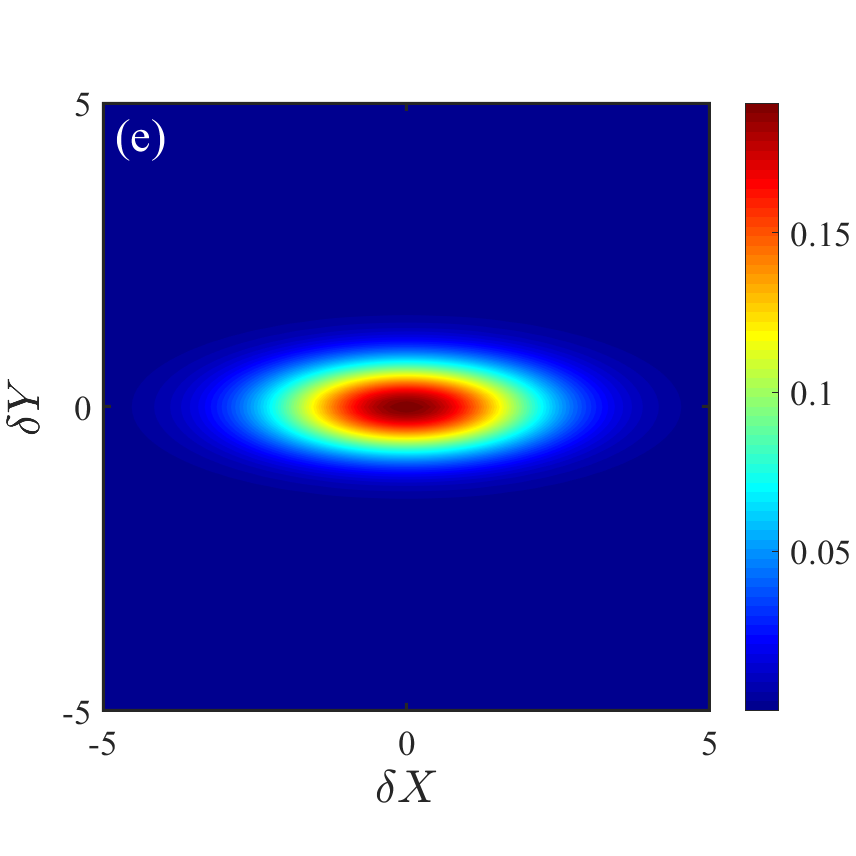}
\includegraphics[width=0.3\columnwidth]{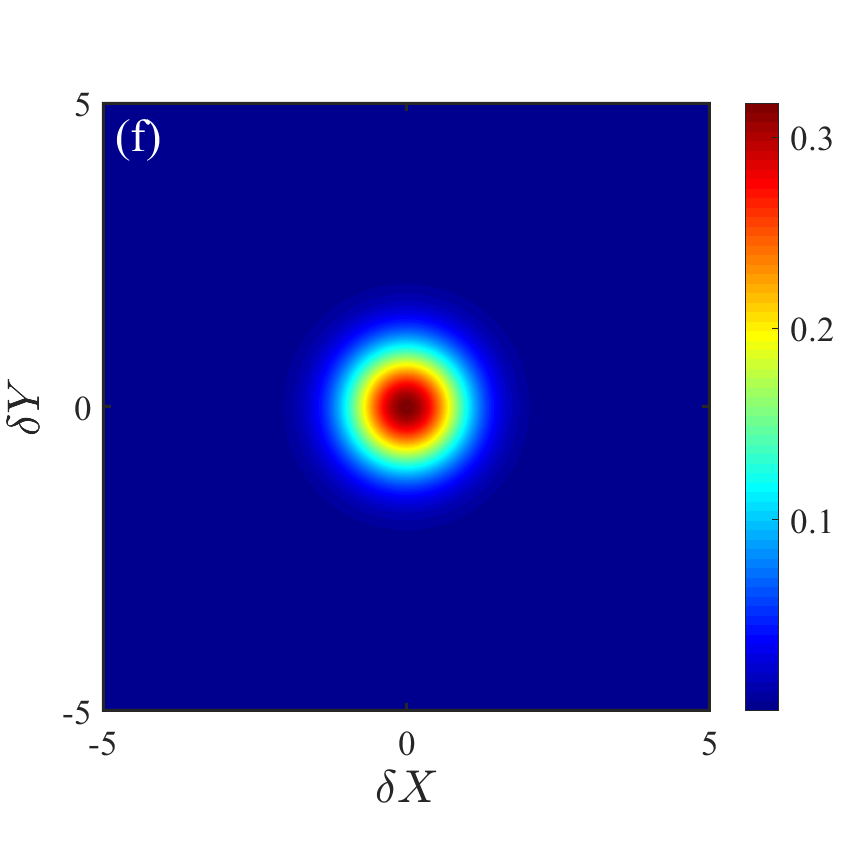}
\caption{The Wigner functions in the phase space. Panels (a) and (b) are the Wigner functions of the mechanical oscillator, and the parameter sets $\{g_+/g_-,G\}$ are $\{0.28,0\}$ and $\{0.28,0.4\kappa\}$, respectively. Panels (c), (d), (e), and (f) illustrate the Wigner functions of the cavity field, and the parameter sets $\{g_+/g_-,G\}$ are chosen as $\{0.28,0.4\kappa\}$, $\{0.28,0\}$, $\{0.91,0.4\kappa\}$ and $\{0.91,0\}$, respectively. The other parameters are the same as those in Fig. \ref{fig:2}.}\label{fig:4}
\end{figure}

The squeezing phenomenon can be visualized through the Wigner function in the phase space. For a mechanical oscillator, its covariance matrix can be written as a $2\times2$ matrix form with
\begin{align}
\mathcal{V}_b= \left[
  \begin{array}{cc}
    \mathcal{V}_{33}  & \mathcal{V}_{34}\\
    \mathcal{V}_{43}  & \mathcal{V}_{44}\\
\end{array}\right],\label{eq:Vb}
\end{align}
and the corresponding Wigner function of the mechanical oscillator can be expressed as \cite{Weedbrook2012}
\begin{align}
\mathcal{W}(\mathcal{R})=&\frac{\exp(-\frac{1}{2}\mathcal{R}^T\mathcal{V}_b^{-1}\mathcal{R})}
{2\pi\sqrt{\det(\mathcal{V}_b)}},\label{eq:W}
\end{align}
where $\mathcal{R}=[Q,P]^T$ represents a two-dimensional vector of the covariance matrix $\mathcal{V}_b$. In Fig. \ref{fig:4}, we vividly display the Wigner functions of the mechanical oscillator and the cavity field in the phase space. Figs. \ref{fig:4}(a) and \ref{fig:4}(b) show the Wigner functions of the mechanical oscillator under the synergistic influence of two-tone driving and parametric pumping, where the parameter sets $\{g_+/g_-,G/\kappa\}$ corresponded to (a) and (b) are $\{0.28,0\}$ and $\{0.28,0.4\}$, respectively. By comparing these two panels, we can observe that the Wigner function is  more severely compressed in horizontal axis when the gain coefficient is $G=0.4\kappa$, which indicates that the joint mechanism of two-tone driving and parametric pumping can effectively enhance the mechanical squeezing. Through Figs. \ref{fig:4}(c)-\ref{fig:4}(f), we can explain the squeezing transfer from the cavity field to the mechanical oscillator by examining the variation in the Wigner functions, where the parameter sets $\{g_+/g_-,G/\kappa\}$ corresponded to (c)-(f) are $\{0.28,0.4\}$, $\{0.28,0\}$, $\{0.91,0.4\}$ and $\{0.91,0\}$, respectively. According to panels (c) and (d), it is evident that the cavity field is not squeezed when $G=0$ under the identical blue-detuned driving condition. Comparing the panels (c) and (e), the vertical axis of the cavity field is compressed  by the similar amount when the component of the blue-detuned driving is increased and the gain coefficient is still set to $G=0.4\kappa$. The result suggests that the degree of squeezing of the cavity field remains almost unchanged. According to panels (e) and (f), it is clear to conclude that even at the optimal ratio of $g_+/g_-$, the shape of the Wigner function of the cavity field does not change in either horizontal or vertical axis when the parametric pumping is absent, indicating that the cavity field is not squeezed. Regardless of how the blue-detuned laser varies, the squeezing of the cavity field depends solely on the gain coefficient $G$, and can be effectively transferred to the mechanical oscillator.

The physical mechanism underlying the generation of mechanical squeezing can be elucidated as follows. Based on the linearized quantum Langevin equations (\ref{eq:youxiaozhangluo}) for the slow-varying operators, the effective interaction Hamiltonian between the cavity field and the mechanical oscillator can be derived in the form of $-\delta\tilde{a}^\dag(g_+\delta\tilde{b}^\dag+g_-\delta\tilde{b}) -\delta\tilde{a}(g_+\delta\tilde{b}+g_-\delta\tilde{b}^\dag)$. By employing the standard squeezing transformation and introducing the Bogoliubov mode $\delta B=\cosh r\delta\tilde{b}+\sinh r\delta\tilde{b}^\dag$, the effective Hamiltonian can, in principle, be rewritten as $-g_{\text{eff}}(\delta\tilde{a}^\dag \delta B+\delta\tilde{a}\delta B^\dag)$, where $g_{\text{eff}}=\sqrt{g_-^2-g_+^2}$ denotes the effective coupling strength between the cavity field and the Bogoliubov mode. The squeezing coefficient $r$ in the Bogoliubov mode is determined by $r=\ln[(g_- +g_+)/(g_- -g_+)]/2$, which is positively correlated with the squeezing of the mechanical oscillator. The squeezing of the cavity field can be transferred to the Bogoliubov mode through a beam-splitter interaction, and a strong effective coupling strength $g_{\text{eff}}$ will intuitively facilitate the squeezing of the mechanical oscillator. The value of $g_{\text{eff}}$ will increase as the $g_+/g_-$ ratio decreases; however, it inevitably results in a reduction of the squeezing coefficient $r$. Therefore, the optimal mechanical squeezing occurs at the point determined by the competitive interaction between $g_{\text{eff}}$ and $r$. Upon analyzing the expression for the squeezing coefficient $r$, it should be noted that the value of the squeezing coefficient is $r=0$ when the component of blue-detuned laser is absent, i.e., $g_+=0$, and that there is no mechanical squeezing when the parametric pumping does not exist. When parametric pumping is injected, the squeezing generated in the cavity field can be transferred to the mechanical oscillator, allowing for strong mechanical squeezing to be realized even when the amplitude ratio of the red/blue-detuned laser is far from optimal range.

\begin{figure}[t]
\centering
\includegraphics[width=0.3\columnwidth]{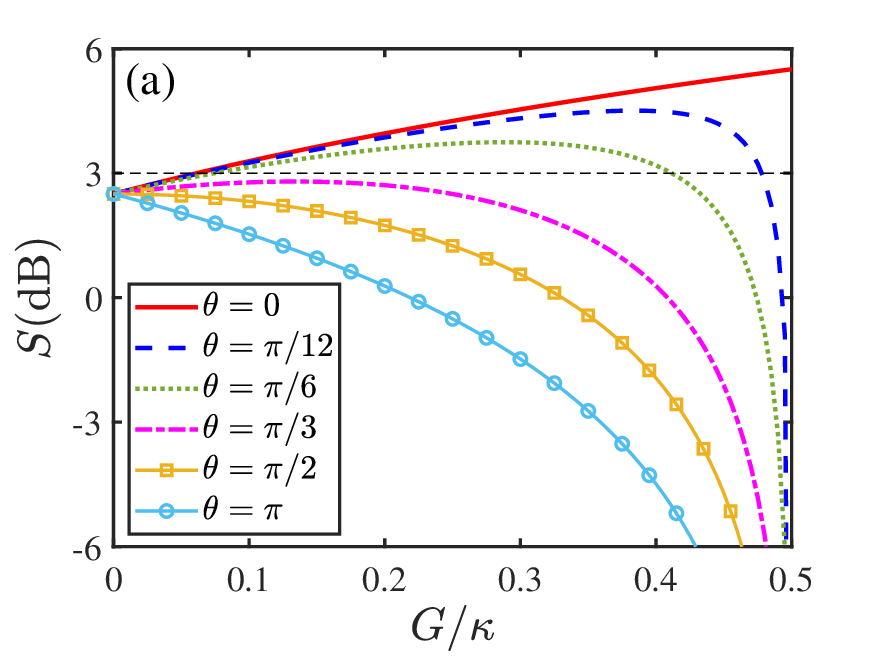}
\includegraphics[width=0.3\columnwidth]{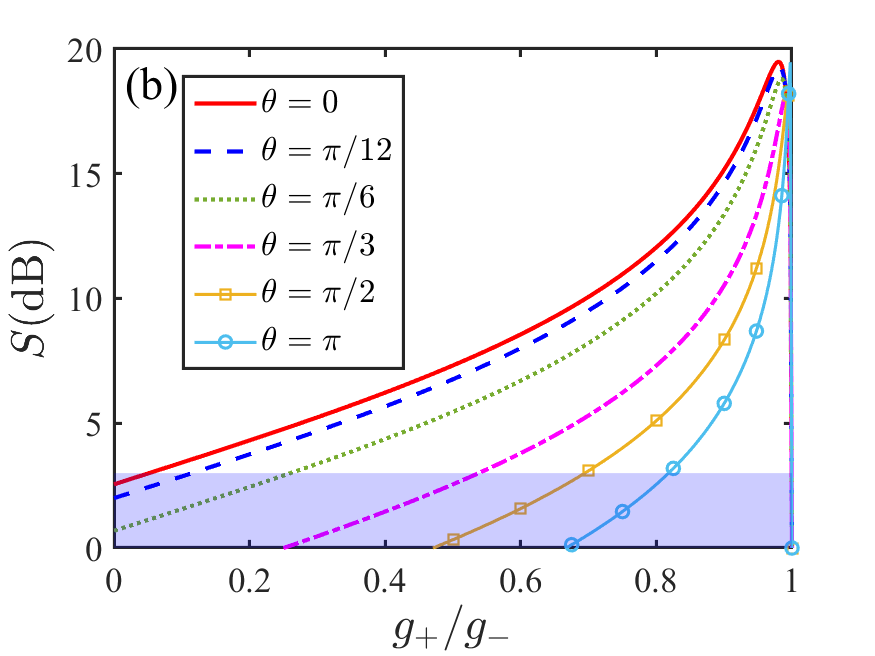}
\includegraphics[width=0.3\columnwidth]{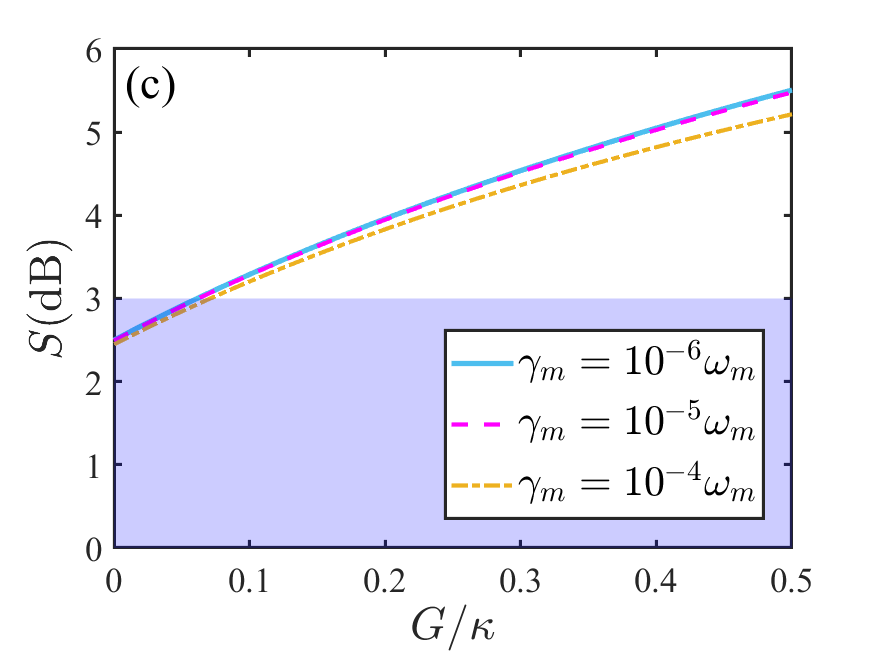}
\includegraphics[width=0.3\columnwidth]{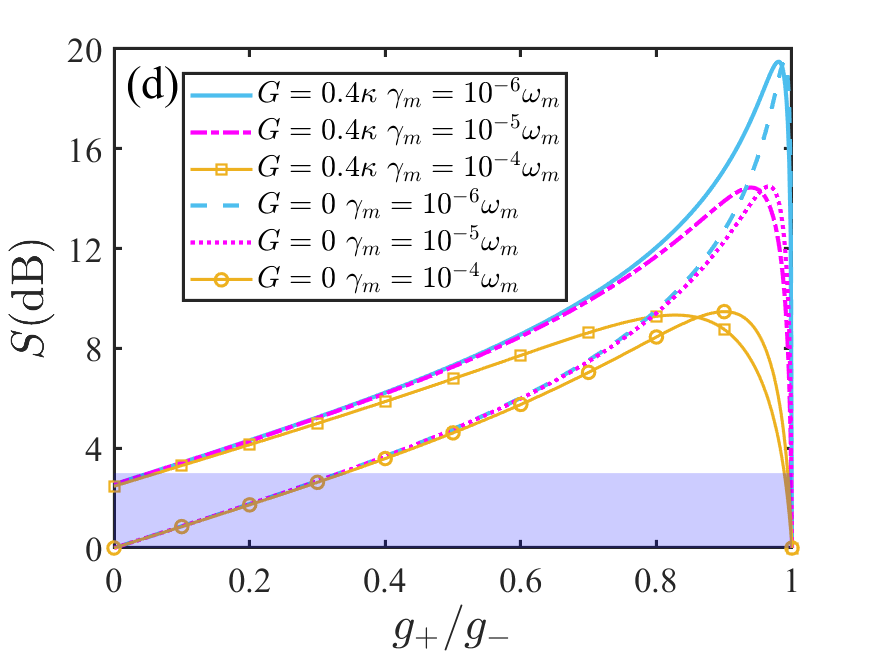}
\includegraphics[width=0.3\columnwidth]{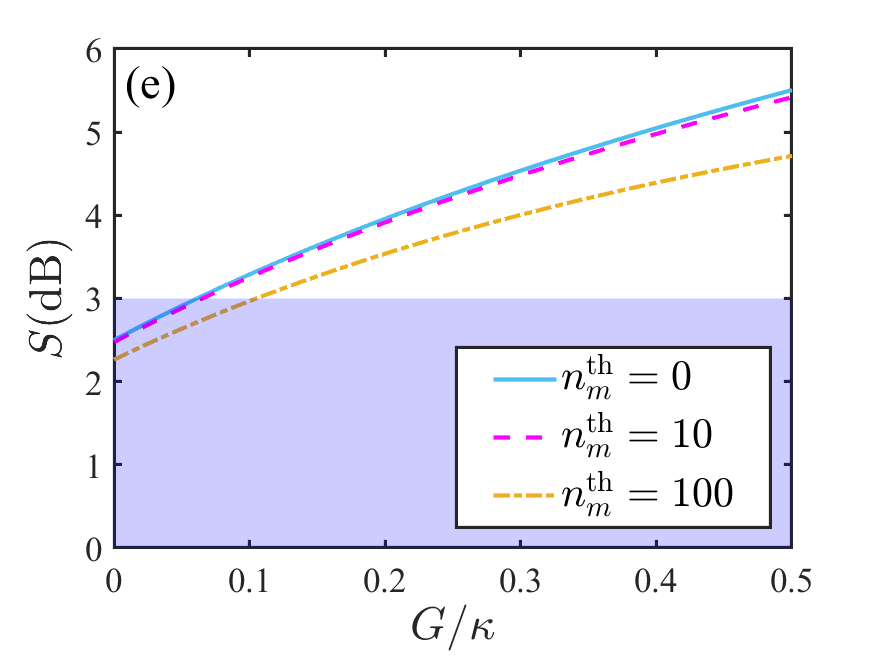}
\includegraphics[width=0.3\columnwidth]{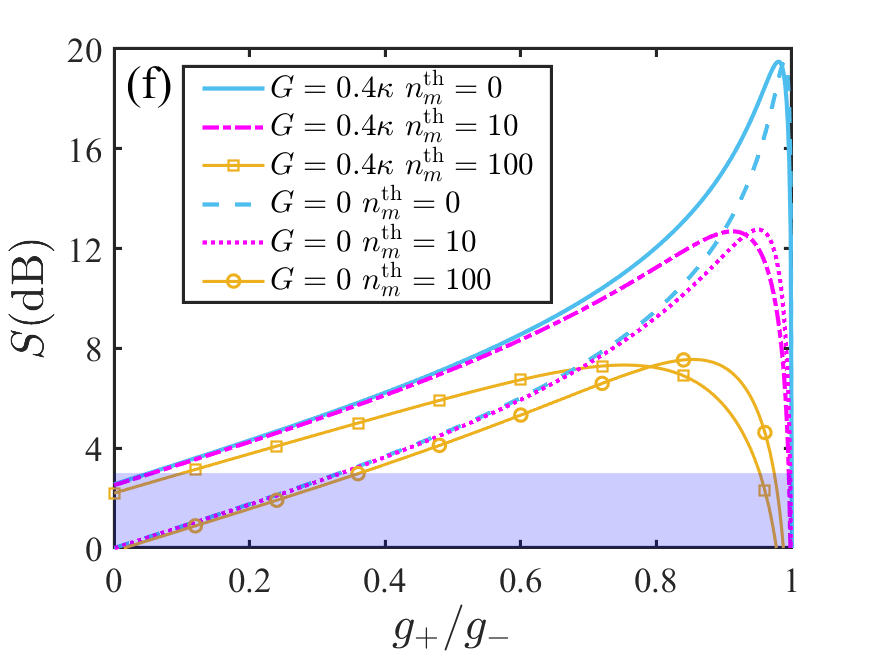}
\caption{The degree of mechanical squeezing as a function of the gain coefficient $G$ (a) and the $g_+/g_-$ ratio (b) are under different relative phases $\theta$. The degree of mechanical squeezing with respect to the gain coefficient $G$ (c) and the $g_+/g_-$ ratio (d) are shown under different decay rates $\gamma_m/\omega_m=\{10^{-4},10^{-5},10^{-6}\}$, respectively. Panels (e) and (f) plot the degree of squeezing as a function of $G$ and $g_+/g_-$ under thermal phonon number $n_m^{\text{th}}=\{0,10,100\}$, respectively. The purple shaded areas represent the regions where the degree of squeezing is below 3 dB. All other parameters are in accordance with those specified in Fig. \ref{fig:2}.}\label{fig:5}
\end{figure}

In Fig. \ref{fig:5}, we investigate how the degree of squeezing of the mechanical oscillator varies with respect to other system parameters, including the relative phase $\theta$, the mechanical oscillator decay rate $\gamma_m$, and the thermal phonon number $n_m^{\text{th}}$. Fig. \ref{fig:5}(a) illustrates the effects of variations in the gain coefficient $G$ on the degree of squeezing when the relative phase changes from $0$ to $\pi$; similarly, Fig. \ref{fig:5}(b) depicts the behavior of the degree of squeezing as a function of the $g_+/g_-$ ratio under the same conditions. The degree of squeezing exhibits a periodic variation of $2\pi$ with respect to $\theta$, and it is optimal for the mechanical squeezing when the relative phase $\theta$ is set to $0$. Furthermore, the degree of squeezing increases monotonically as the gain coefficient $G$ increases, indicating that the phase of the parametric pumping $\theta=0$ is ideal, which results in maximum efficiency during the process of generating mechanical squeezing. Generally speaking, as the $g_+/g_-$ ratio increases, the degree of mechanical squeezing will also increase. However, this trend reaches a maximum when the $g_+/g_-$ ratio approaches $1$. When the parametric pumping is executed, the achievement of mechanical squeezing beyond the 3 dB limit becomes feasible across almost the entire ratio range of $g_+/g_-$. In Fig. \ref{fig:5}(c), the mechanical squeezing modulations with the gain coefficient $G$ are shown under mechanical decay rates $\gamma_m/\omega_m=\{10^{-4}, 10^{-5}, 10^{-6}\}$, respectively. With different mechanical decay rates, Fig. \ref{fig:5}(d) reveals the degree of squeezing as a function of the $g_+/g_-$ ratio, capturing not only the squeezing behavior in the absence of parametric pumping but also the condition  when parametric pumping is employed. These two panels demonstrate the subtle influence of the decay rate $\gamma_m$ on the mechanical squeezing, which has a potential to significantly reduce the requirements for the quality of micro/nano fabrication techniques. Even with high mechanical decay rate, $\gamma_m=10^{-4}\omega_m$, a strong mechanical squeezing can still be achieved. However, the 3 dB squeezing limit cannot be broken in the range that is dominated by the red-detuned laser when the parametric pumping is not employed. Except for the case where the laser is almost the red-detuned or the amplitudes of the red/blue-detuned laser are almost equal, a strong squeezing beyond the 3 dB limit can be achieved in most ratio range of $g_+/g_-$ when the parametric pumping is added, such as, $G=0.4\kappa$. In Figs. \ref{fig:5}(e) and \ref{fig:5}(f), we analyze the impact of thermal phonon number $n_m^{\text{th}}$ on the degree of squeezing. Even the thermal phonon number reaches to $n_m^{\text{th}}=100$, the mechanical squeezing beyond the 3 dB limit is still observed, which demonstrates that the proposed scheme is robust against thermal noise. What's more, the adverse effects induced by the surrounding environment can be suppressed by the joint using of two-tone driving with appropriate $g_+/g_-$ ratio and parametric pumping.

\section{Analytical solution}\label{sec5}
In this section, we will present an analytical solution for the degree of squeezing (\ref{eq:S}), and subsequently analyze the joint effect of two-tone driving and parametric pumping. Under the weak coupling limit, i.e., $g_+,g_-\ll\kappa$, the motion of the mechanical oscillator is significantly slower than the variations of the optical cavity field. As a result, the cavity field can be regarded as an instantaneous response to the motion of the mechanical oscillator, rather than being dynamically coupled to it; therefore, the cavity field and the mechanical oscillator can be considered as effectively decoupled or adiabatically interacted with each other \cite{Aspelmeyer2014}. In our model, the position of the mechanical oscillator can be considered a slowly varying parameter, whereas the cavity field rapidly adjusts itself to accommodate these slow changes. When the system reaches a steady state, we can set $\delta\dot{\tilde{a}}=0$ and obtain that
\begin{align}
\delta\tilde{a}=\frac{1}{\kappa^2-4G^2}\{i[(2\kappa g_+ -4Ge^{i\theta}g_-)\delta\tilde{b}^\dag+(2\kappa g_- -4Ge^{i\theta}g_+)\delta\tilde{b}]+4Ge^{i\theta}\sqrt{\kappa}\tilde{a}^\dag_{\text{in}}
+2\kappa\sqrt{\kappa}\tilde{a}_{\text{in}}\}.
\label{eq:delta_ tilde_a}
\end{align}
By substituting $\delta\tilde{a}$ into Eq. (\ref{eq:youxiaozhangluo}), $\delta\dot{\tilde{b}}$ can be rearranged as
\begin{align}
\delta\dot{\tilde{b}}=&A\delta\tilde{b}+B\delta\tilde{b}^\dag+C\tilde{a}_{\text{in}}(t)+D\tilde{a}^\dag_{\text{in}}(t)
+E\tilde{b}_{\text{in}}(t).
\label{eq:delta_dot_tildeb}
\end{align}
Here, the coefficients are given by
\begin{align}
A=&\frac{-g_-}{\kappa^2-4G^2}(2\kappa g_--4Ge^{i\theta}g_+)+\frac{g_+}{\kappa^2-4G^2}(2\kappa g_+-4Ge^{-i\theta}g_-),\notag\\
B=&\frac{-g_-}{\kappa^2-4G^2}(2\kappa g_+-4Ge^{i\theta}g_-)+\frac{g_+}{\kappa^2-4G^2}(2\kappa g_--4Ge^{-i\theta}g_+),\notag\\
C=&\frac{i\sqrt{\kappa}}{\kappa^2-4G^2}(4Ge^{-i\theta}g_+ +2\kappa g_-),\notag\\
D=&\frac{i\sqrt{\kappa}}{\kappa^2-4G^2}(4Ge^{i\theta}g_- +2\kappa g_+),\notag\\
E=&\sqrt{\gamma_m}.
\label{eq:A_B_C_D_E}
\end{align}
In Eq. (\ref{eq:delta_dot_tildeb}), the term dependent on $\gamma_m$ has been neglected. As a result, we can subsequently obtain the dynamical equation for the position fluctuation $\delta Q$ of the mechanical resonator as
\begin{align}
\delta\dot{Q}=&\frac{2(g_+^2-g_-^2)}{\kappa+2G}\delta Q+F_1(t)+F_2(t),
\label{eq:delta_dot_Q}
\end{align}
where,
\begin{align}
F_1(t)=&\frac{i\sqrt{2\kappa}(g_- -g_+)}{\kappa+2G}[\tilde{a}_{\text{in}}(t)-\tilde{a}^\dag_{\text{in}}(t)],\notag\\
F_2(t)=&\sqrt{\frac{\gamma_m}{2}}[\tilde{b}_{\text{in}}(t)+\tilde{b}^\dag_{\text{in}}(t)].
\label{eq:F1_F2}
\end{align}
The correlation functions of $F_1(t)$ and $F_2(t)$ satisfy the following relationship
\begin{align}
\langle F_1(t)F_1(t^{\prime})\rangle=&\frac{2\kappa(g_- -g_+)^2}{(\kappa+2G)^2}\delta(t-t^{\prime}),\notag\\
\langle F_2(t)F_2(t^{\prime})\rangle=&\frac{\gamma_m}{2}(2n_m^{\text{th}}+1)\delta(t-t^{\prime}).
\label{eq:guanlianhanshu}
\end{align}

\begin{figure}[t]
\centering
\includegraphics[width=0.45\columnwidth]{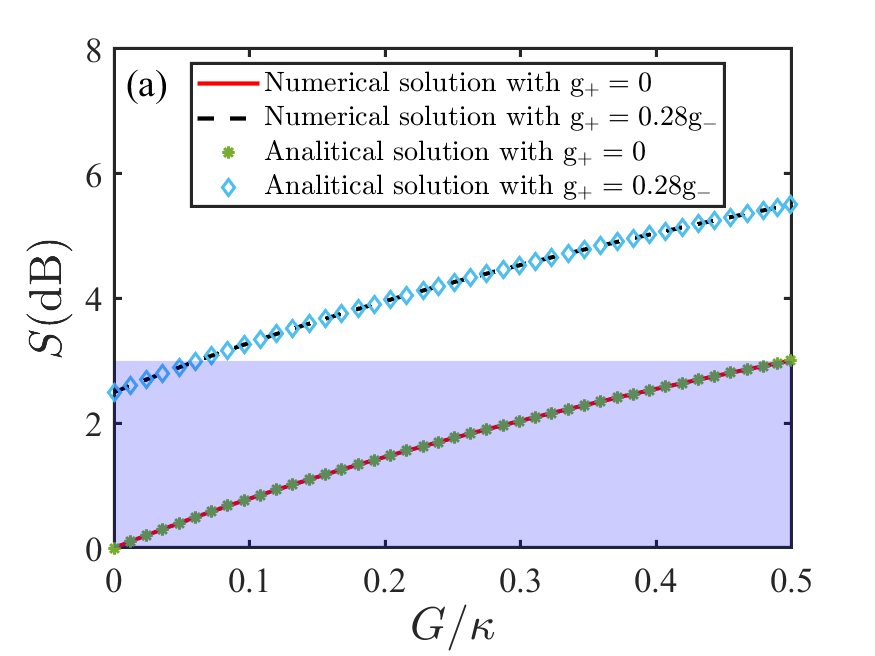}
\includegraphics[width=0.45\columnwidth]{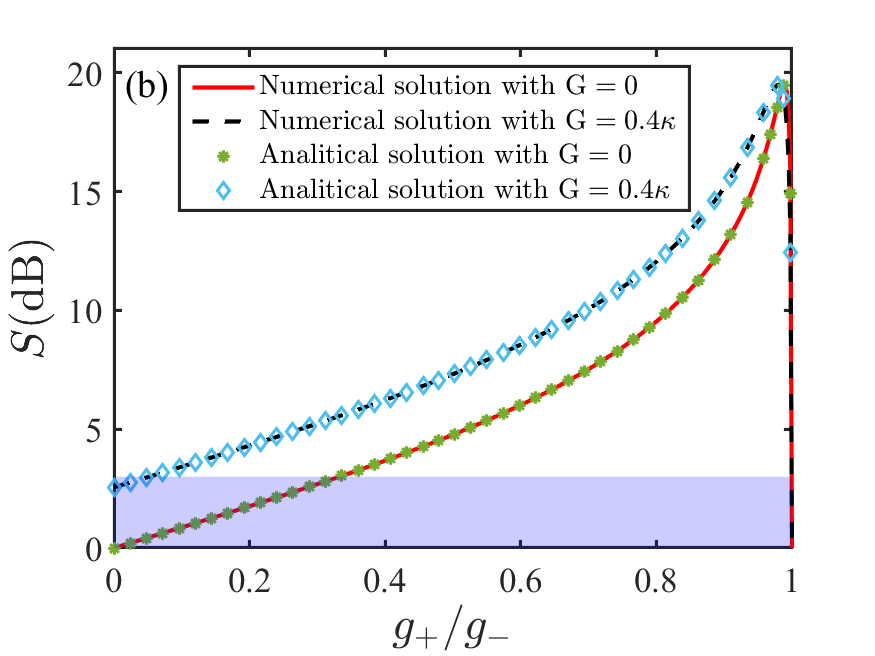}
\caption{The degree of mechanical squeezing is plotted using both numerical solution (red solid line and black dashed line) and analytical solution (Eq. (\ref{eq:S_jiexi}), green stars and cyan diamonds), respectively. The parameters are the same as those in Fig. \ref{fig:2}. }\label{fig:6}
\end{figure}

From Eqs. (\ref{eq:delta_dot_Q}-\ref{eq:guanlianhanshu}), we can derive the dynamical equation for the fluctuation of the position operator as
\begin{align}
\frac{d\langle\delta Q^2\rangle}{dt}=&\frac{2\kappa(g_- -g_+)^2}{(\kappa+2G)^2}+\frac{\gamma_m}{2}(2n_m^{\text{th}}+1)
+\frac{4(g_+^2 -g_-^2)}{\kappa+2G}\langle\delta Q^2\rangle.
\label{eq:dot_deltaQ_jiexi}
\end{align}
Thus, we can calculate the analytical solution for the variance of the position operator as
\begin{align}
\langle\delta Q^2\rangle=&\frac{\kappa(g_--g_+)}{2(\kappa+2G)(g_- +g_+)}
+\frac{\gamma_m(\kappa+2G)}{8(g_-^2 -g_+^2)}(2n_m^{\text{th}}+1).
\label{eq:deltaQ_jiexi}
\end{align}

In the steady-state case, the degree of squeezing of the position operator, which is expressed in units of dB, can be provided by the following analytical form:
\begin{align}
S=&-10\log_{10}\left[\frac{\kappa(g_- -g_+)}{2(\kappa+2G)(g_- +g_+)} +\frac{\gamma_m(\kappa+2G)}{8(g_-^2 -g_+^2)}\right]-10\log_{10}2,\label{eq:S_jiexi}
\end{align}
where the thermal phonon number $n_m^{\text{th}}$ is assumed as $0$. In Fig. \ref{fig:6}, we present a comparison of the numerical and analytical results for the mechanical oscillator's degree of squeezing. Fig. \ref{fig:6}(a) depicts the degree of squeezing in relation to the gain coefficient $G$, whereas Fig. \ref{fig:6}(b) illustrates the degree of squeezing that correlates with the $g_+/g_-$ ratio. Through the competition between these two methods, it is evident that the analytical and numerical solutions are in excellent agreement, irrespective of the choice of the independent variable.

\section{Conclusion}\label{sec6}
In conclusion, we have proposed a straightforward scheme that harnesses the joint mechanism of two-tone driving and parametric pumping to achieve and generate robust mechanical squeezing, surpassing the 3 dB threshold. Our finding reveals that the interaction between the cavity field and the mechanical oscillator can be effectively modulated through careful selection of the parametric pumping frequency. Moreover, the squeezing initially generated in the cavity field by the degenerate OPA can be further transferred to the mechanical oscillator, which has already undergone squeezing due to the two-tone driving. This transfer mechanism enhances the mechanical squeezing compared to the scenario where only one mechanism is applied. The scheme is capable of achieving strong mechanical squeezing even when the amplitude ratio of the two-tone driving lasers is far from optimal, where such squeezing would otherwise be unattainable without the application of parametric pumping. We have also obtained the analytical solution for the degree of squeezing, which is in excellent agreement with the numerical results. The generated mechanical squeezing exhibits robustness against thermal noise and mechanical decay, potentially reducing the requirements for device quality. This innovative approach not only significantly enhances our understanding of the fundamental mechanisms of light-matter interaction but also opens up new avenues for potential applications in the field of optomechanical system.

\begin{backmatter}
\bmsection{Funding}
National Natural Science Foundation of China (Grant No. 12204440), Fundamental Research Program of Shanxi Province (Grant Nos. 20210302123063 and 202103021223184), Shanxi Province Billion Project (Grant Nos. 11010538001, 11013410, 11010538020 and 11010538027).

\bmsection{Disclosures}
The authors declare no conflicts of interest.

\bmsection{Data Availability}
The data that support the findings of this study are available upon reasonable request from the authors.
\end{backmatter}

\newcommand{\doi}[2]{\href{https://doi.org/#1}{\color{blue}#2}}

\end{document}